\documentclass[pra,aps,twocolumn]{revtex4}
\usepackage[12pt]{xy}
\usepackage{graphics}
\usepackage{color}

\begin{document}

\title{Witnessing Entanglement with Second-Order Interference}

\author{Magdalena Stobi\'nska} \email{magda.stobinska@fuw.edu.pl}
\affiliation{Instytut Fizyki Teoretycznej, Uniwersytet Warszawski,
  Warszawa 00--681, Poland} \author{Krzysztof W\'odkiewicz}
\email{wodkiew@fuw.edu.pl} \affiliation{Instytut Fizyki Teoretycznej,
  Uniwersytet Warszawski, Warszawa 00--681, Poland}

\affiliation{Department of Physics and Astronomy, University of New
  Mexico, Albuquerque, NM~87131-1156, USA}

\date{\today}

\begin{abstract}
  Second-order interference and Hanbury-Brown and Twiss type
  experiments can provide an operational framework for the
  construction of witness operators that can test classical and
  nonclassical properties of a Gaussian squeezed state (GSS), and
  provide entanglement witness operators to study the separability
  properties of correlated Gaussian squeezed sates.
\end{abstract}

\pacs{????}

\maketitle

\section{Introduction}

Two-party systems described by Gaussian states play an important role
in quantum information \cite{Branstein2002}, continuous-variables (CV)
teleportation \cite{Furusawa1998} and CV entanglement
\cite{Heersink2003}. Such states provide physical realizations of
entangled CV resources, required to transmit quantum information in
highly efficient quantum channels. The general theory of quantum
separability for quantum Gaussian squeezed states has been established
recently in several papers \cite{Englert2002}.

In most theoretical applications squeezed two-mode Gaussian states
\cite{Schumaker1986} have been used as a physical realization of the
original Einstein Podolsky Rosen (EPR) correlated wave function
\cite{Ou1992}.  Such Gaussian squeezed states can be generated in
nonlinear optical parametric amplifiers or in four-wave-mixing
processes \cite{Yurke1983}. Recently an efficient method used the Kerr
nonlinearity in optical fibers to generate squeezed entanglement in
various variables, like phase and amplitude \cite{Silberhorn2001},
polarization \cite{Korolkova2002} and photon-number
\cite{Schmitt1998}.

This experimental progress has stimulated various theoretical tools of
diagnostics of CV states which can provide a measure of entanglement
and squeezing of such states. So far we know that entanglement of such
CV sates can be tested using the Heisenberg uncertainty relations
\cite{Duan2002}, the partial positivity transposition (PPT) criterion
\cite{Simon2000}, the interference and particle counting
\cite{Cirac2003}, or using quantum witness operators
\cite{Horodecki1996}.

In this paper, we propose to employ the Hanbury-Brown and Twiss (HBT)
interference to witness the quantum nature of one-mode squeezed states
and the quantum entanglement of two-mode Gaussian squeezed states.
This method seems to be rather easy experimentally realizable
measuring second-order correlation function of light. We show that
using HBT interference one can witness: classical and nonclassical
properties of single mode Gaussian squeezed states and test quantum
features of two-mode uncorrelated squeezed states. With the help of
HBT witness operator we study the entanglement of a general EPR states
obtained as a nonlocal action of the beam splitter on two input
squeezed states. This witness operator is closely related to the HBT
interference.

This paper is organized as follows: section II is devoted to a review
of well known properties of one-mode Gaussian squeezed state (GSS).
The fundamental properties of such state are derived.  The
$P$-representability condition is analyzed using a single mode quantum
witness operator. Relations with a recent experiment are discussed.
Section III introduces HBT interference and discusses the classical
versus quantum properties of the second-order visibility. A witnessing
operator to detect quantum features of this second-order interference
is introduced.  Quantum properties of two uncorrelated squeezed modes
in terms of interference visibility are investigated. In section IV,
HBT entanglement witnessing is analyzed for an unsqueezed EPR mixed
state. This witnessing operator becomes useful in the discussion of CV
Werner states.  In Section V we discuss entanglement generated by the
action of a beam splitter on two squeezed states. For such general
states the properties of the HBT interference are discussed.  Finally
some conclusions are presented.

\section{One-Mode Gaussian Squeezed State}

We start our discussion with a brief review of a one-mode GSS of the
radiation field. As it is known \cite{Englert2003,Gardiner}, the most
general expression describing the density operator of a GSS
characterized by creation and annihilation operators $(a,a^{\dagger})$
has the form of a Gaussian operator. Up to linear shifts, this
operator can be written in the following normally ordered form
\begin{equation}
\rho = \frac{1}{\sqrt{D}}:\exp{\big[-\frac{(n+1)}{D}a^{\dagger}a -
\frac{m}{2D}\, a^{\dagger\, 2}-
  \frac{m^*}{2D}\, a^{2}\big]}:\,, \label{eq:rho_sec2}
\end{equation}
with $D=(n+1)^2-|m|^2$, where the two parameters, $n$ and
$m=|m|e^{i\lambda}$, are related to the following field correlations
\begin{equation}
\langle a^{\dagger}a \rangle = n,\; \langle aa\rangle = -m,\; \langle
a^{\dagger} a^{\dagger}\rangle = -m^*.
\end{equation}
In order to represent a physical state $(\rho \ge 0)$ these parameters
have to satisfy the following inequality
\begin{equation}
|m|^2 \le n(n+1)\,,\label{eq:positivity_criterion_sec2}
\end{equation}
where the equality holds only if the density operator is a pure state
$\rho=|\Psi\rangle\langle \Psi|$.

Various phase-space distributions, like the Wigner function or the
diagonal $P$-representation of the density operator
(\ref{eq:rho_sec2}) can be evaluated taking the Fourier transform of
the Weyl characteristic function
\begin{equation}
C(\alpha) = \mathrm{Tr}\{D(\alpha) \rho \} = e^{-\frac{1}{2}
  (\alpha^*, \, \alpha) \mathbf{C} {\alpha \choose \alpha^*}},
\end{equation}
where $D(\alpha)$ is the displacement operator, $\rho$ is the density
operator given by (\ref{eq:rho_sec2}) and
\begin{eqnarray}
\mathbf{C} = \left( \begin{array}{cc}
n+\frac{1}{2} & m  \\
m^* & n+\frac{1}{2}
\end{array} \right)\label{eq:matrixC_sec2}
\end{eqnarray}
is the covariance matrix of these Gaussian states.

In general, a positive Gaussian operator is said to be
$P$-representable if there exists its decomposition in a coherent
projector basis~\cite{Scully1997}
\begin{equation}
\rho = \int \mathrm{d^2} \alpha \, P(\alpha) |\alpha \rangle \langle
\alpha|,
\end{equation}
with a non-singular phase-space distribution function $P(\alpha)$.
The GSS state given by (\ref{eq:rho_sec2}) is $P$-representable if and
only if $n>|m|$, and in this case we have
\begin{equation}
P(\alpha)= \frac{1}{\pi\sqrt{d}}\,\exp{\big[-\frac{n}{d}|\alpha|^2
- \frac{m}{2d}\, \alpha^{\ast\, 2}-
  \frac{m^*}{2d}\, \alpha^{2}\big]}\,,
\end{equation}
where $d=n^2-|m|^2$. This Gaussian distribution with $m=0$,
corresponds to a thermal state with a mean number of photons given by
$n$. A $P$-representable squeezed Gaussian state can be called
classical because its statistical properties can be simulated from the
above classical Gaussian distribution function of the field amplitude
and phase.

\begin{figure}[h]
\begin{center}
\scalebox{0.8}{\includegraphics{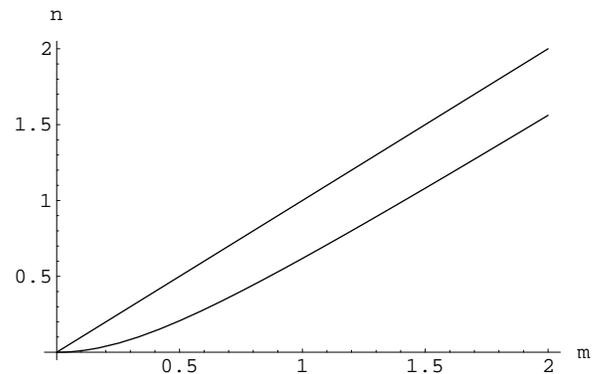}}
\caption{Quantum and classical  one-mode GSS given by
  (\ref{eq:matrixC_sec2}).  According to constraint
  (\ref{eq:positivity_criterion_sec2}), values on or above the curve
  specify physical states. Values between the curve and the line
  describe squeezed, not $P$-representable, quantum states.
  $P$-representable states belong to values on or above the line
  $n=|m|$.} \label{Fig:1_sec2}
\end{center}
\end{figure}

In Fig.~\ref{Fig:1_sec2} we have depicted values of ($n$, $m$) for
which the mixed Gaussian state is classical ($P$-representable) or
nonclassical (is not $P$-representable).

Continuous-variable states of light can be described by their electric
field amplitude and phase quadratures
\begin{equation}
X_1 = \frac{a + a^{\dagger}}{\sqrt{2}}, \;
X_2 = \frac{a - a^{\dagger}}{\sqrt{2}i},
\end{equation}
\noindent
with the commutator $[X_1,X_2]=i$.

The uncertainties of these two quadratures are given by the following
relations
\begin{eqnarray}
\left({\Delta X_1}\right)^2 &=& \frac{ \langle a^{\dagger}
a\rangle + \langle a
  a^{\dagger} \rangle + \langle a^2 \rangle + \langle
  {a^{\dagger}}^2 \rangle}{2} \nonumber \\
&=& n + \frac{1}{2} -|m|\cos\lambda\,,\\
\left({\Delta X_2}\right)^2 &=& \frac{\langle a^{\dagger} a\rangle
+
 \langle a a^{\dagger} \rangle -\langle a^2 \rangle - \langle
  {a^{\dagger}}^2 \rangle}{2} \nonumber \\
 &=& n + \frac{1}{2} + |m|\cos\lambda\,.
\end{eqnarray}

From these uncertainty relations we see that a classical GSS can
reduce thermal fluctuations (characterized by the mean number of
photons $n$), to the vacuum level.  Quantum GSS can reduce the quantum
fluctuations in one quadrature (e.g.  amplitude) below the vacuum
level at the expense of increased fluctuations in the conjugate
variable (phase)
\begin{equation}
\left({\Delta X_1}\right)^2 < \frac{1}{2}.
\end{equation}
Such a reduction cannot be described by a $P$-representable GSS.

As it is well known, the nonclassical nature of GSS can be revealed in
a second-order degree of coherence, even though the photon statistics
of the vacuum squeezed state is super-Poissonian~\cite{Scully1997}.
For a general mixed GSS, the normalized second-order degree of
coherence is
\begin{equation}
g^{(2)} = \frac{\langle a^{\dagger} a^{\dagger}aa\rangle}{\langle
  a^{\dagger} a \rangle^2} = 2 + \frac{|m|^2}{n^2},
\end{equation}
where $g^{(2)}=3$ is the border value between classical and
nonclassical GSS. States, for which $ 3\leq g^{(2)} \le
3+\frac{1}{n}$, are quantum, so not $P$-representable. Note that for a
small mean number of photons ($n < 1$), $g^{(2)}$ can be very large.

The nonclassical properties of a GSS, can be witnessed using the
following quantum witness operator
\begin{equation}
\mathcal{W}^{(2)}=3-\frac{
  a^{\dagger} a^{\dagger}aa}{\langle a^{\dagger} a \rangle^2}\,.
\end{equation}
In this case
  \begin{equation}
\mathrm{Tr}\{  \mathcal{W}^{(2)}\rho\}= \frac{n^2 -|m|^2}{n^2}\,
\end{equation}
is positive for $P$-representable GSS, and negative for not
$P$-representable GSS. Pure state is always quantum, because
\begin{equation}
\langle \Psi| \mathcal{W}^{(2)}|\Psi \rangle= -\frac{1}{n}\,.
\end{equation}

A GSS with controllable parameters $n$ and $m$ has been recently
generated in an optical parametric amplifier with a phase-insensitive
amplifier of gain $H$ followed by a phase-sensitive amplifier of gains
$G$ and $1/G$ \cite{Wenger2004}. In such a generator the incoming mode
$a_{in}$ is transformed as follows
\begin{eqnarray}
a &=& \frac{1}{2}\Big( \frac{1}{\sqrt{G}} + \sqrt{G}\Big) \sqrt{H} \;
a_{in} +
\frac{1}{2}\Big(\frac{1}{\sqrt{G}} - \sqrt{G}\Big) \sqrt{H} \;
a^{\dagger}_{in} \nonumber \\
&+&  \frac{1}{2}\Big( \frac{1}{\sqrt{G}} - \sqrt{G}\Big) \sqrt{H-1} \;
v + \frac{1}{2}\Big( \frac{1}{\sqrt{G}} + \sqrt{G}\Big) \sqrt{H-1} \;
v^{\dagger}, \nonumber \\
\end{eqnarray}
where $v$, $v^{\dagger}$ are the amplifier quantum noise boson
operators. Assuming that modes $a_{in}$ and $v$ are initially empty we
obtain that the outgoing state is a GSS with
\begin{eqnarray}
n &=& \frac{1}{2}\Big( \frac{1}{G} + G\Big)\Big(H- \frac{1}{2}\Big) -
\frac{1}{2}\,, \nonumber \\
|m| &=& \frac{1}{2}\Big( G -\frac{1}{G}\Big)\Big(H- \frac{1}{2}\Big).
\end{eqnarray}
Such a state is $P$-representable (classical) if $H \ge
\frac{1}{2}(G+1)$, and the purity of this state depends only on the
phase-insensitive gain $Tr\{\rho^2\} = \frac{1}{2H-1}$.

Changing the pump average power in the experiment, states with $1 \le
G_{exp} \le 1.65$ and $1 \le H_{exp} \le 1.05$ have been obtained. In
this range the GSS was quantum, i.e. not $P$-representable. For the
maximum value of $G_{exp} \simeq 1.65$ and $H_{exp} \simeq 1.05$ the
value of $H$ is $20\%$ below the classical threshold. The
corresponding value of the normalized second-order degree of
coherence, for these gain parameters is purely quantum, $g^{(2)} =
7.68$, with $n \simeq 0.12$ and $|m| \simeq 0.29$, leading to
$\mathrm{Tr}\{ \mathcal{W}^{(2)}\rho\}=- 4.84$, and a purity close to
$91\%$.

\section{Second-order Interference as a Quantum Witness}

Considering first-order correlation functions and Young-type
experiments is not enough to distinguish classical and quantum
properties of GSS.  Moreover, there is no first-order interference for
thermal state at all. It is well known that a second-order
interference and Hanbury-Brown and Twiss type
experiments~\cite{Scully1997}, can reveal quantum properties of light
sources, not seen in Young experiments.

\begin{figure}[h]
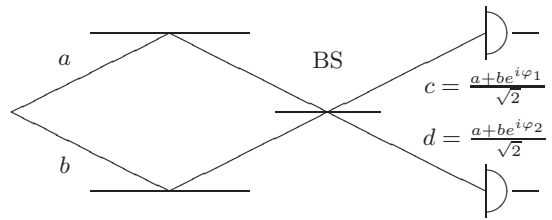

\begin{center}
  \begin{displaymath}
    \xy<0.7cm,0cm>:
    (-1.5,1.5);(1.5,1.5)**@{-},
    (-1.5,-1.5);(1.5,-1.5)**@{-},
    (2,0);(4,0)**@{-}, (3,1)*{\mbox{BS}},
    (-3,0);(0,1.5)**@{-};(6,-1.5)**@{-},
    (-3,0);(0,-1.5)**@{-};(6,1.5)**@{-},
    (-2,1)*{a}, (-2,-1)*{b},
    (6,2);(6,1)**@{-}, (6,1.5)*\cir(0.5,0.5){r^l},
    (6.5,1.5);(7,1.5)**@{-},
    (6,0.5)*{c=\frac{a+b  e^{i\varphi_1}}{\sqrt{2}}},
    (6,-2);(6,-1)**@{-}, (6,-1.5)*\cir(0.5,0.5){r^l},
    (6.5,-1.5);(7,-1.5)**@{-},
    (6,-0.5)*{d=\frac{a+b e^{i\varphi_2}}{\sqrt{2}}},
   \endxy
  \end{displaymath}
\end{center}
\caption{Hanbury-Brown and Twiss second-order interference.}
\label{Fig:2_sec3}
\end{figure}

A HBT experiment that we have in mind is depicted on
Fig.~\ref{Fig:2_sec3}. The source produces two light beams with
quantum amplitudes $a$ and $b$. The interference pattern at the screen
is proportional to $1+v\cos(\varphi_1-\varphi_2)$, where $\varphi_1$
and $\varphi_2$ are phase differences between the beams $a$ and $b$ in
front of the detectors. These phases include geometrical phases and
phases due to the possible action of the beam splitter. The
second-order visibility $v$ of these fringes can reveal the quantum
nature of the light source. For a classical source the quantum
amplitudes of the field are replaced by classical amplitudes and the
classical visibility of the fringes $v_{\mathrm{cl}}$ is always either
equal or less than $50\%$. This classical limit is violated for single
photons, as it has been shown in the pioneering experiments performed
by Mandel, reviewed in \cite{Mandel1999}.

Let us investigate the HBT interference resulting from two independent
GSSs represented by modes $(a,b)$, for which the density operator is
separable (uncorrelated) and has the form
\begin{equation}
\rho = \rho(a) \otimes \rho(b)\,.
\end{equation}

At the detector the positive-frequency part of electric field
(normalized to the number of photons) is as follows
\begin{equation}
E^{(+)}(\varphi_i) = \frac{a+be^{i\varphi_i}}{\sqrt{2}}\,.
\end{equation}

The field intensity operator at the screen is equal to
\begin{equation}
I(\varphi_i)=E^{(-)}(\varphi_i)E^{(+)}(\varphi_i)=\frac{1}{2}( I_a
+I_b+ b^\dagger a\,e^{-i\varphi_i} +
    a^\dagger b\,e^{i\varphi_i})\,,
\end{equation}
where $I_a= a^\dagger a$ and $I_b= b^\dagger b$.

From this expression, we obtain that the normally ordered second-order
intensity correlation is \begingroup \setlength{\arraycolsep}{0pt}
\begin{eqnarray}
\label{eq:HBT_correlation_sec3}
 \langle{:}I(\varphi_1)\,I(\varphi_2){:}\rangle
&=& \frac{1}{4}\Big[
\langle :(I_a+I_b)^2: \rangle + 2\langle :I_a I_b:\rangle
\cos(\varphi_1-\varphi_2) \nonumber \\
&+& (e^{-i\varphi_1} + e^{-i\varphi_2})\langle
b^{\dagger}:(I_a+I_b):a\rangle \nonumber \\
&+& (e^{i\varphi_1} + e^{i\varphi_2})\langle
a^{\dagger}:(I_a+I_b):b\rangle \\
&+& e^{-i(\varphi_1+\varphi_2)}\langle {b^{\dagger}}^2a^2  \rangle
+ e^{i(\varphi_1+\varphi_2)}\langle {a^{\dagger}}^2 b^2 \rangle
\Big]. \nonumber
\end{eqnarray}
\endgroup

If we assume that the two modes have the same mean number of photons
$\langle a^{\dagger}a\rangle = \langle b^{\dagger}b\rangle = n$ and
the squeezing parameters $\langle a^2\rangle = -m$ and $\langle
b^2\rangle = -m e^{i\lambda}$ differ only by a phase $\lambda$, we
obtain the intensity correlation function in the form
\begin{equation}
\label{eq:HBT_correlation2_sec3}
 \langle{:}I(\varphi_1)\,I(\varphi_2){:}\rangle \propto 1 + v_{-}
 \cos(\varphi_1 - \varphi_2) +
v_{+}\cos(\varphi_1 + \varphi_2 + \lambda)\,,
\end{equation}
where
\begin{eqnarray}
v_{-} = \frac{2\langle : I_a I_b :\rangle}{\langle :(I_a+I_b)^2:
  \rangle}, && \; v_{+} = \frac{\langle {b^{\dagger}}^2 a^2  \rangle + \langle
{a^{\dagger}}^2 b^2 \rangle}{\langle :(I_a+I_b)^2 :\rangle}, \nonumber\\
v_{-} = \frac{n^2}{3n^2 + |m|^2}, && \; v_{+} = \frac{|m|^2}{3n^2 +
 |m|^2}\,. \label{eq:visibilities_sec3}
\end{eqnarray}

In these calculations, normally ordered correlation functions like
$\langle \left.{a}^{\dagger}\right.^2 \left.{a}^2\right.  \rangle$,
$\langle \left.{a}^{\dagger}\right.^2 ab\rangle$ and similar, have
been evaluated using the Gaussian decomposition of fourth-order
correlations into second-order correlations. This follows from the
fact that for Gaussian states only second moments are needed to
calculate higher moments of the field correlations.

The formulas
(\ref{eq:HBT_correlation_sec3}-\ref{eq:HBT_correlation2_sec3}), are
different from similar formulas involving second-order interference of
single photons. Note that the interference pattern consists of two
terms being in-phase and out-of-phase $\varphi_{1} \mp \varphi_{2}$.
For unsqueezed states the out-of-phase term vanishes.

The interference pattern presented in this equation is genuine to
squeezed light. The in-phase visibility $v_{-}$ appears always, no
matter whether the two interfering states are squeezed or not. The
out-of-phase visibility $v_{+}$, proportional to $|m|^2$ is entirely
due to the squeezing of the incoming modes.  If in a realistic
experimental situation, the relative phase $\lambda$ between the two
squeezed modes can be controlled, the out-of-phase term contributes to
the interference pattern. Note that selecting $\varphi_1-\varphi_2 =
\frac{\pi}{2}$, the interference pattern consists of only out-of-phase
fringes proportional to $v_{+}$, and the actual position of the maxima
and minima depends on the phase~$\lambda$.

For an unsqueezed source of light, $|m|=0$, the interference pattern
is thermal with a visibility $v_{-} = \frac{1}{3}$. At the border line
between classical and quantum squeezing, $|m|=n$, the two visibilities
are equal $v_{-}=v_{+}= 25\%$. For quantum GSS we always have $v_{-}
\le v_{+}$. From these relations we see that the pure GSS is always
quantum. Only mixed GSS can reach the classical region. The classical
behavior of GSS seen in second-order interference, results in the
following set of classical inequalities
\begin{equation}
\frac{1}{4} \le v_{-} \le \frac{1}{3}, \; \; 0 \le v_{+} \le
\frac{1}{4}, \; \; v_{-} + v_{+} \le \frac{1}{2}\,.
\label{eq:classical_inequalities_sec3}
\end{equation}

\begin{figure}[h!]
\begin{center}
  \scalebox{0.8}{\includegraphics{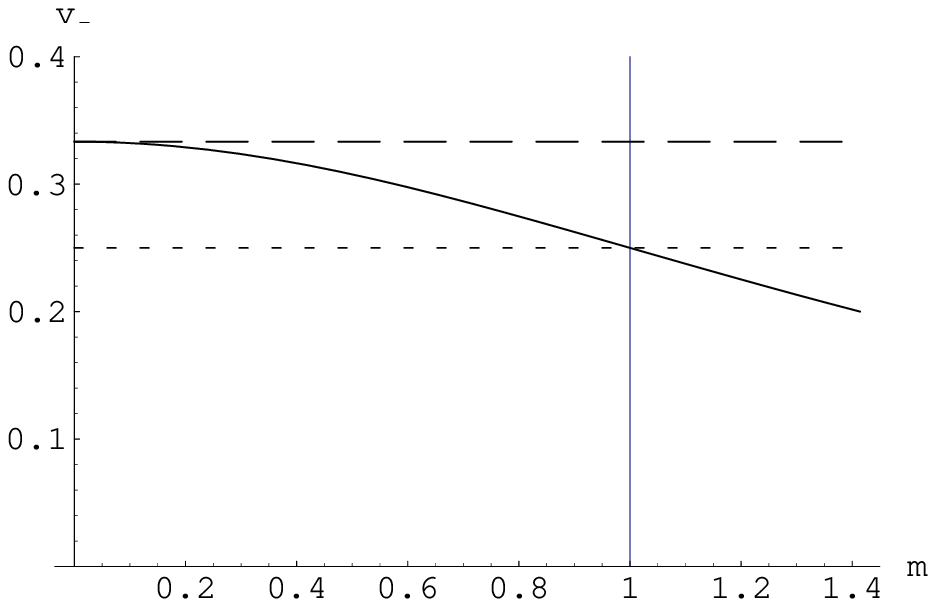}}
  \scalebox{0.8}{\includegraphics{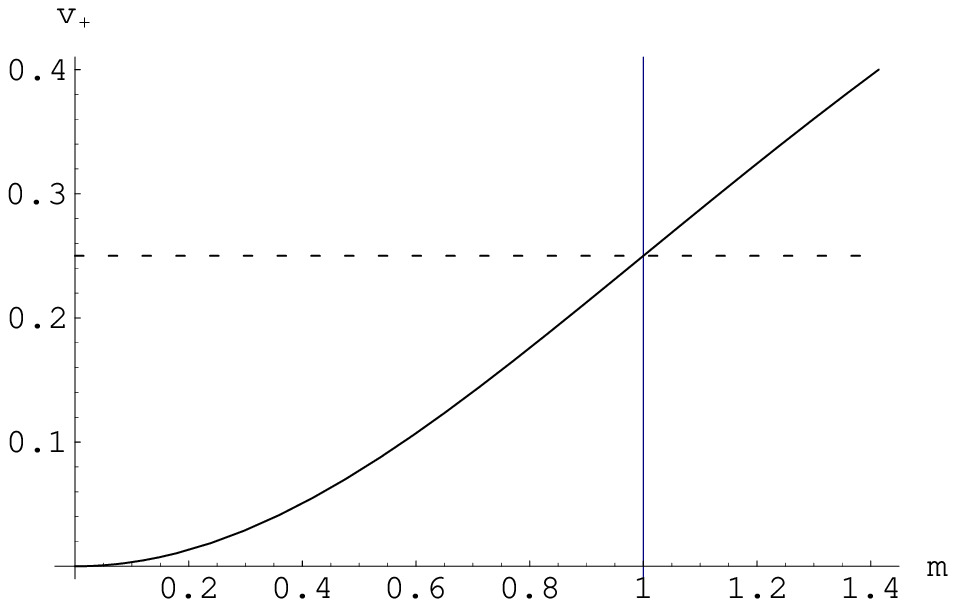}}
\caption{The visibilities $v_{-}$ and $v_{+}$
  (\ref{eq:visibilities_sec3}) evaluated for $n=1$.  The vertical grid
  line defines a border between classical GSS (values of $|m|<1$) and
  quantum GSS ($|m|>1$). The visibility $v_-$ is always bounded from
  above by $\frac{1}{3}$ --- the dashed horizontal grid line.  The
  dotted grid line, $\frac{1}{4}$, separates the values of $v_-$ and
  $v_+$ corresponding to classical and to quantum GSS.  $v_-$ is
  bounded from above and $v_+$ is bounded from below by $\frac{1}{4}$
  for quantum GSS.  The value of $|m|=\sqrt{2}$ corresponds to the
  pure state.}
\label{Fig:3_sec3}
\end{center}
\end{figure}

\begin{figure}[h!]
\begin{center}
  \scalebox{0.8}{\includegraphics{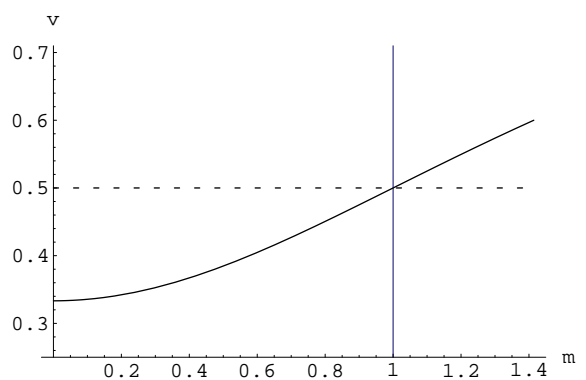}}
\caption{The sum $v_{-}+v_{+}$ evaluated for $n=1$ is depicted.
  For quantum GSS ($|m|>1$ --- vertical grid line) the sum is always
  bounded from below by $\frac{1}{2}$ --- horizontal grid line.}
\label{Fig:4_sec3}
\end{center}
\end{figure}

For quantum sources of GSS, these inequalities are violated, as it is
shown in Fig.~\ref{Fig:3_sec3} and Fig.~\ref{Fig:4_sec3}, both plotted
for two-mode GSS with fixed $n=1$. The line $m=1$ is the border
between quantum and classical states. The values of $m$ between $1$
and $\sqrt{2}$ correspond to quantum GSS, $n<|m|$. The $m=\sqrt{2}$ is
the maximal value of this parameter which obeys the positivity
constraint $|m|^2 = n(n+1)$. It corresponds to the pure state.  The
maximum second-order visibility $v = v_- + v_+$ is achieved for angles
$\varphi_1=\varphi_2$ and $\lambda = \varphi_1+\varphi_2$.

We conclude this section, introducing a HBT witness operator
\begin{equation}
\label{eq:witness_sec3}
 \mathcal{W}^{(HBT)}=\frac{1}{2}- \frac{2 a^\dagger  b^\dagger a b
 + {b^{\dagger}}^2 a^2 + {a^{\dagger}}^2
  b^2 }{\langle :(I_a+I_b)^2:
  \rangle}\,.
\end{equation}
For this witness operator we have
  \begin{equation}
\mathrm{Tr}\{  \mathcal{W}^{(HBT)}\rho\}=  \frac{n^2
-|m|^2}{2(3n^2 + |m|^2)}\,,
\end{equation}
which is positive for $P$-representable GSS, and negative for non
$P$-representable GSS.

For the estimated experimental values of the squeezed state discussed
in the previous sections, we obtain: $v_{-} \simeq 11\%$, $ v_{+}
\simeq 66\%$ and $v \simeq 77\%$, resulting with the the value:
$\mathrm{Tr}\{ \mathcal{W}^{(HBT)}\rho\} \simeq -0.274$, i.e. well
below the classical limit.

We will see that the second-order witness (\ref{eq:witness_sec3}) can
be applied, with some precautions, as an entanglement witness for a
correlated squeezed states.

\section{Interference Witness of Mixed EPR State}

Let us investigate a two-mode Gaussian correlated state of Alice and
Bob given by a density operator $\rho_{\mathrm{AB}}$. The entangled
resource is ideally a two-mode GSS \cite{Schumaker1986} which has
annihilation operators $a$ and $b$. It is well known that such state
can be generated in a process of nonlinear optical parametric
amplification. For this two-mode Gaussian correlated state the
covariance matrix of the Weyl characteristic function, $C(\alpha,
\beta) = \mathrm{Tr}\{ \rho_{\mathrm{AB}} D(\alpha) D(\beta) \}$, is a
$4 \times 4$ matrix with the following matrix elements
\begin{eqnarray}
\mathbf{C}_{\mathrm{AB}} = \left( \begin{array}{cccc}
   n+\frac{1}{2} & 0 & 0 & m_c \\
   0 & n+\frac{1}{2} & m_c^* & 0 \\
   0 & m_c & n+\frac{1}{2} & 0 \\
   m_c^* & 0 & 0 &  n+\frac{1}{2}
\end{array} \right), \label{eq:matrixC_sec4}
\end{eqnarray}
where $\langle a^{\dagger}a \rangle = \langle b^{\dagger}b \rangle =
n$ and $\langle ab \rangle = -m_c$ is the correlation parameter
between the two modes.

In general, in accordance with Werner's separability
criterion~\cite{Werner1989}, the two-mode quantum density operator is
said to be separable if it is a sum of product states
\begin{equation}
\rho_{\mathrm{AB}} = \sum_{i} p_i \, \rho^{i}(a) \otimes
\rho^{i}(b),
\end{equation}
where $\rho^{i}(a)$, $\rho^{i}(b)$ are the density operators of the
two modes and $\sum_{i} p_i =1$, $p_i > 0$.

As it has been discussed in several papers (see the tutorial
\cite{Englert2003} and references therein), this two-mode state is the
optical analog of the entangled EPR wave function if $|m_c|^2 =
n(n+1)$ and $n \to \infty$. For $n<|m_c|$ the correlated state is
entangled, so it is nonseparable and, in general, mixed.

For the mixed EPR state described above the HBT intensity correlation
function is a sum of mean values of the following terms
\begin{eqnarray}
\langle{:}I(\varphi_1)\,I(\varphi_2){:}\rangle = \frac{1}{4}
\langle \left.{{a}^{\dagger}}\right.^2 {a}^2
\rangle + \frac{1}{4} \langle \left.{b}^{\dagger}\right.^2
\left.{b}^2\right. \rangle  \nonumber\\
+ \frac{1}{2} \langle  \left.{a}^{\dagger}{b}^{\dagger}
ab\right.\rangle (1+\cos(\varphi_1 - \varphi_2)),
\end{eqnarray}
which can be reduced to
\begin{equation}
\langle{:}I(\varphi_1)\,I(\varphi_2){:}\rangle =
\frac{1}{2}(3n^2 + |m_c|^2)
\left(1+  v_{-} \cos(\varphi_1 -\varphi_2) \right),
\end{equation}
where  the visibility is
\begin{equation}
v_{-} = \frac{n^2 + |m_c|^2}{3n^2 + |m_c|^2}\,. \label{eq:visibility_sec4}
\end{equation}

In case of no correlation, i.e. if $|m_c| = 0$, we get $v_{-} =
\frac{1}{3}$ as for a thermal state. For separable states we get
$\frac{1}{3} \leq v_- \leq \frac{1}{2}$. For some values of parameters
$n$ and $m_c$ the second-order visibility violates the classical limit
of $\frac{1}{2}$. For nonseparable, i.e. entangled states, the
visibility is always quantum: $v_{-}> \frac{1}{2}$.

\begin{figure}[h!]
\begin{center}
  \scalebox{0.8}{\includegraphics{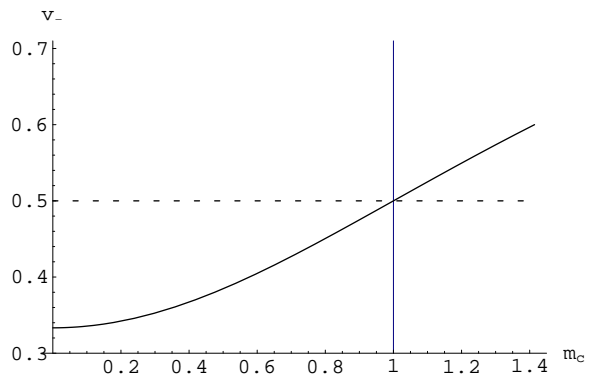}}
\caption{The visibility $v_{-}$ (\ref{eq:visibility_sec4}) of
  the EPR state. The vertical grid line defines the border between
  values of $|m_c|$ which correspond to separable and to entangled
  states.  For $|m_c|>1$ the state is entangled and the visibility is
  greater than $\frac{1}{2}$ --- horizontal grid line.}
\label{Fig:5_sec4}
\end{center}
\end{figure}

In Fig.~\ref{Fig:5_sec4} we have depicted this visibility for a mixed
EPR states with $n=1$, as a function of $m_c$.  The range of the
correlation parameter, $1\le |m_c| \le \sqrt{2}$, corresponds to mixed
nonseparable states, with the exception of the upper boundary,
corresponding to a pure state. In this case the maximal visibility is
$v_-=0.6$.

For this mixed two-mode EPR state, we can use the quantum witness
operator given by (\ref{eq:witness_sec3}) as an entanglement witness.
It is easy to see that the last two terms of the witness operator
related to single mode squeezing do not contribute and as a result we
obtain
\begin{equation}
\mathrm{Tr}\{  \mathcal{W}^{(HBT)}\rho_{\mathrm{AB}}\}=  \frac{n^2
-|m_c|^2}{2(3n^2 + |m_c|^2)}\,,
\end{equation}
which is positive for separable mixed EPR states and negative for non
nonseparable mixed EPR states.

HBT interference is also useful when detecting entanglement in a
Werner-like state. A natural extension of a Werner state for infinite
dimensions is a convex combination of a correlated state with the
separable maximally mixed states. In this case the Werner state is a
combination of a mixed EPR state and two uncorrelated thermal states
\begin{equation}
\label{eq:werner_sec4}
 \rho_{W} = p \rho_{AB} + (1-p)\rho^{\mathrm{T}}_{A}
\otimes \rho^{\mathrm{T}}_{A},
\end{equation}
where $0 \le p \le 1$.  The mean number of photons in each mode $A$
and $B$ is the same and equal to $n$. Simple calculation shows that
the visibility of this state is given by
\begin{equation}
v_- = \frac{n^2 + p |m_c|^2}{3n^2 + p |m_c|^2}\,.
\end{equation}

From this calculation we see that effectively the Werner state
corresponds to a mixed EPR state with correlation parameter $m_c$
scaled by $\sqrt{p}$. The visibility of this state exceeds 50\%, i.e.
the entanglement witness (\ref{eq:witness_sec3}) turns negative if
$n^2 \leq p|m_c|^2$. If the EPR state is a maximally entangled state
with $|m_c|^2=n(n+1)$, this condition simplifies and reads
\begin{equation}
p > \frac{n}{n+1}\,.\label{eq:HBTcriterion_sec4}
\end{equation}

For the Werner state with a maximally entangled EPR state, the
sufficient criterion for separability involving the positive partial
transposition (PPT) can be used and has the form \cite{Wenger2004}
\begin{equation}
p > \left(1+\sqrt{\frac{1+n}{n}} \,
  \frac{(1+2n^2)^2}{n(1+2n)(1+n^2)} \right)^{-1}. \label{eq:PPT_sec4}
\end{equation}

\begin{figure}[h!]
\begin{center}
  \scalebox{0.8}{\includegraphics{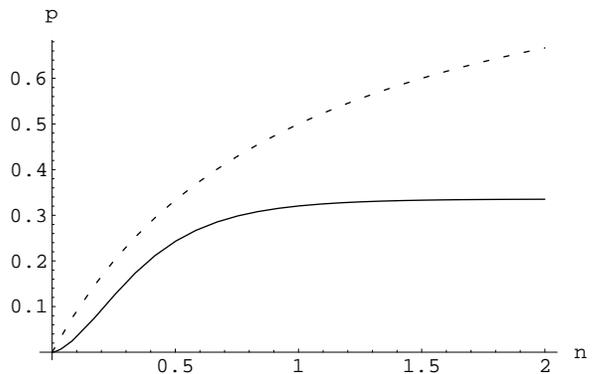}}
\caption{Comparison of two separability criterions: the dotted line
  corresponds to HBT interference criterion
  (\ref{eq:HBTcriterion_sec4}) and the solid line to PPT criterion
  (\ref{eq:PPT_sec4}). According to criterions states above the curves
  are nonseparable.}
\label{Fig:6_sec4}
\end{center}
\end{figure}

As it is depicted in Fig.~\ref{Fig:6_sec4}, the criterion based on the
HBT entanglement witness is stronger than the PPT criterion.  For the
linear combination forming the continuous-variable Werner state given
by (\ref{eq:werner_sec4}), no proof of the separability condition
exists.

\section{HBT interference of entangled squeezed states}

\begin{figure}[h]
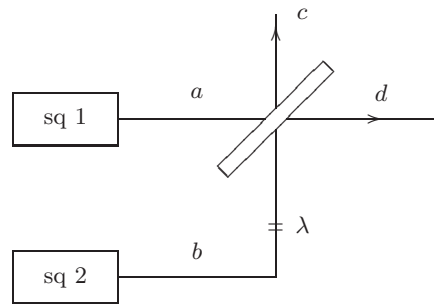

\begin{center}
\begin{displaymath}
    \xy<0.7cm,0cm>:
    (-5,0.5);(-3,0.5)**@{-};(-3,-0.5)**@{-};(-5,-0.5)**@{-};(-5,0.5)**@{-},
    (-4,0)*\txt{sq 1},
    (-5,-2.5);(-3,-2.5)**@{-};(-3,-3.5)**@{-};(-5,-3.5)**@{-};(-5,-2.5)**@{-},
    (-4,-3)*\txt{sq 2},
    (-3,0);(-0.2,0)**@{-}, (0.2,0);(3,0)**@{-}, (2,0)*\dir{>},
    (-3,-3);(0,-3)**@{-};(0,-0.2)**@{-}, (0,0.2);(0,2)**@{-},
    (0,1.75)*\dir{>}, (0,-2)*\dir{||},
    (-1.5,-2.5)*{b},
    (-1.5,0.5)*{a},
    (2,0.5)*{d},
    (0.5,2)*{c},
    (0.5,-2)*{\lambda},
    (-1.1,-0.9);(0.9,1.1)**@{-},
    (-0.9,-1.1);(1.1,0.9)**@{-},
    (0.9,1.1);(1.1,0.9)**@{-},
    (-0.9,-1.1);(-1.1,-0.9)**@{-},
    \endxy
  \end{displaymath}
\caption{Schematic experimental setup: two, initially separable,
  amplitude squeezed beams, shifted in phase
  (\ref{eq:mean_values_sec5}), interfere at the beam splitter and
  become entangled. The detectors measure second-order correlation
  function between modes $c$ and $d$.}
\label{Fig:7_sec5}
\end{center}
\end{figure}

The EPR state discussed in the previous section can be generated in a
nonlinear optical amplifier. In this section, we discuss entangled
two-mode states resulting from a nonlocal operation on two
uncorrelated single mode GSSs. This efficient way of producing
correlated state uses a beam splitter as the nonlocal operation, that
entangles squeezed states of light. A typical setup involves two
amplitude squeezed beams, initially separable and shifted in phase,
which interfere at the 50/50 beam splitter (Fig.~\ref{Fig:7_sec5}).

The input squeezed states are characterized by the following
parameters
\begin{equation}
\langle a^{\dagger}a \rangle = \langle b^{\dagger}b \rangle =
n, \; \langle a^2 \rangle = \langle b^2 \rangle = -m.
\label{eq:mean_values_sec5}
\end{equation}
We recognize in these relation GSSs from Section II. As it is
indicated in Fig. 7, the squeezed beam $b$ is shifted by a phase
$\lambda$.

The relation between input and output modes operators on the beam
splitter have the well known form
\begin{equation}
c = \frac{a+be^{i\lambda}}{\sqrt{2}}, \;
d = \frac{a-be^{i\lambda}}{\sqrt{2}}.
\end{equation}

As a result of this transformation the outcome is also a GSS with the
covariance matrix of the modes $c$ and $d$ of the form
\begin{eqnarray}
\mathbf{C}_{out} = \left( \begin{array}{cccc}
   n+\frac{1}{2} & \tilde{m} & 0 & m_c \\
  \tilde{m}^{\ast}& n+\frac{1}{2} & m_c^* & 0 \\
   0 & m_c & n+\frac{1}{2} &  \tilde{m} \\
   m_c^* & 0 & \tilde{m}^{\ast} &  n+\frac{1}{2}
\end{array} \right)\,. \label{eq:matrixC_sec5}
\end{eqnarray}

Note that the two output beams form an entangled squeezed EPR state
with equal mean number of photons in each mode $\langle
c^{\dagger}c\rangle = \langle d^{\dagger}d\rangle=n$, equal squeezing
parameters in each mode $\langle cc\rangle = \langle
dd\rangle=-\tilde{m}=-\frac{1}{2}\, m(1 + e^{2i\lambda})$ and with
correlation parameter $\langle cd\rangle =-m_c=- \frac{1}{2}m(1
-e^{2i\lambda})$. We see the crucial role of the angle $\lambda$ in
the correlation parameter.  The correlation parameter $m_c$ is
non-zero valued if $\lambda \neq 0, \pi$ and for other values of
$\lambda$ can lead to a state that is nonseparable. For example if
$\lambda=\frac{\pi}{2}$ the state (\ref{eq:matrixC_sec5}) reduces to
the EPR state (\ref{eq:matrixC_sec4}) with $m_c=m$.

States similar to (\ref{eq:matrixC_sec5}) were produced
in~\cite{Glockl2003} using bright beams.  In this case the
experimental setup consisted of a Sagnac interferometer which produced
two independently squeezed beams. In this experimental paper it has
been assumed that the output state is a pure unsqueezed EPR state
(\ref{eq:matrixC_sec4}).

For the general $\lambda$-dependent state, with the tools developed
for Gaussian states \cite{Englert2003}, it is possible to establish
the following exact condition for separability
\begin{equation}
|m|^2 + \sqrt{\frac{1-\cos 2\lambda}{2}}|m| \le n(n+1)\,,
\label{eq:separability_criterion_sec5}
\end{equation}
or
\begin{equation}
\sqrt{\left(|m|^2 + \sqrt{\frac{1-\cos
2\lambda}{2}}|m|+\frac{1}{4}\right)} \le
n\,. \label{eq:separability_criterion2_sec5}
\end{equation}

\begin{figure}
\begin{center}
  \scalebox{0.8}{\includegraphics{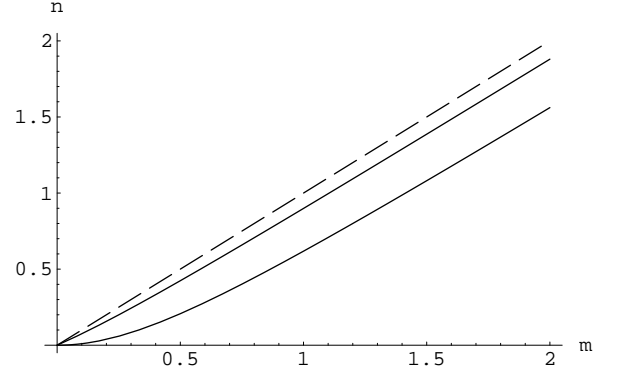}}
  \caption{The constraint (\ref{eq:separability_criterion_sec5}) for
    different values of $\lambda$ is depicted. The line corresponds to
    $\lambda=\frac{\pi}{2}$ --- EPR state (\ref{eq:matrixC_sec4}), the
    upper curve to $\lambda=\frac{\pi}{4}$ --- mixed squeezed EPR
    state and the lower to $\lambda=0$ --- separable state.}
\label{Fig:8_sec5}
\end{center}
\end{figure}

\begin{figure}
\begin{center}
  \scalebox{0.8}{\includegraphics{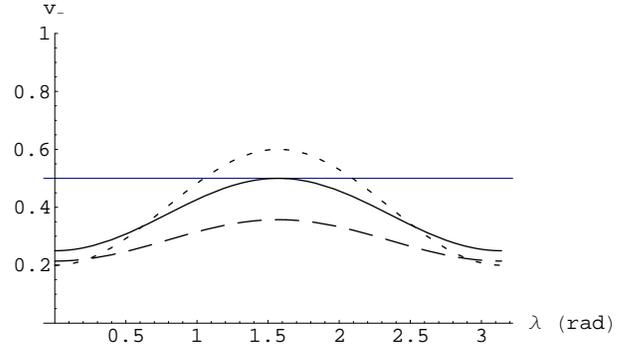}}
  \caption{Depending on the phase $\lambda$ for the quantum input states,
    the output state is either of quantum or classical nature. Its
    $v_-$, the dotted curve, can exceed $\frac{1}{2}$ --- grid
    horizontal line. For input states with $n=|m|$ the maximum value
    of $v_-$ for output state is $\frac{1}{2}$, the solid line.
    Classical input states never give entangled output, the dashed
    curve never reaches the value of $\frac{1}{2}$.}
\label{Fig:8.1_sec5}
\end{center}
\end{figure}

In Fig.~\ref{Fig:8_sec5} we have depicted this condition for three
different angles. Note that for incoming not $P$-representable states,
the outcome produced by the beam splitter is never entangled if
$\cos2\lambda =1$ and is always entangled if $\cos2\lambda =-1$. As an
example, for $\cos2\lambda =0$, there are states that are not
entangled, despite the quantum nature of the incoming beams. To
conclude, it is impossible to produce entangled output from classical
input states.  For quantum input states, depending on $\lambda$, one
can produce either entangled or separable state
(Fig.~\ref{Fig:8.1_sec5}).

The physical significance of the phase $\lambda$ is easy to understand
when considering the quadratures of the input state. It enables
rotation of the error ellipse for a squeezed state in quadrature phase
space. Let us consider the quadrature of the input mode
$be^{i\lambda}$ rotated by an angle $\frac{\theta}{2}$
\begin{equation}
X_{\theta} = \frac{be^{i(\lambda-\frac{\theta}{2})} +
  b^{\dagger}e^{-i(\lambda -\frac{\theta}{2})}}{\sqrt{2}}.
\end{equation}
Assuming $m$ to be real for simplicity, its uncertainty is equal to
\begin{equation}
\langle X_{\theta}^{2} \rangle = n+\frac{1}{2} - m \cos(\theta -
2\lambda).
\end{equation}

\begin{figure}[h!]
\begin{center}
  \scalebox{0.4}{\includegraphics{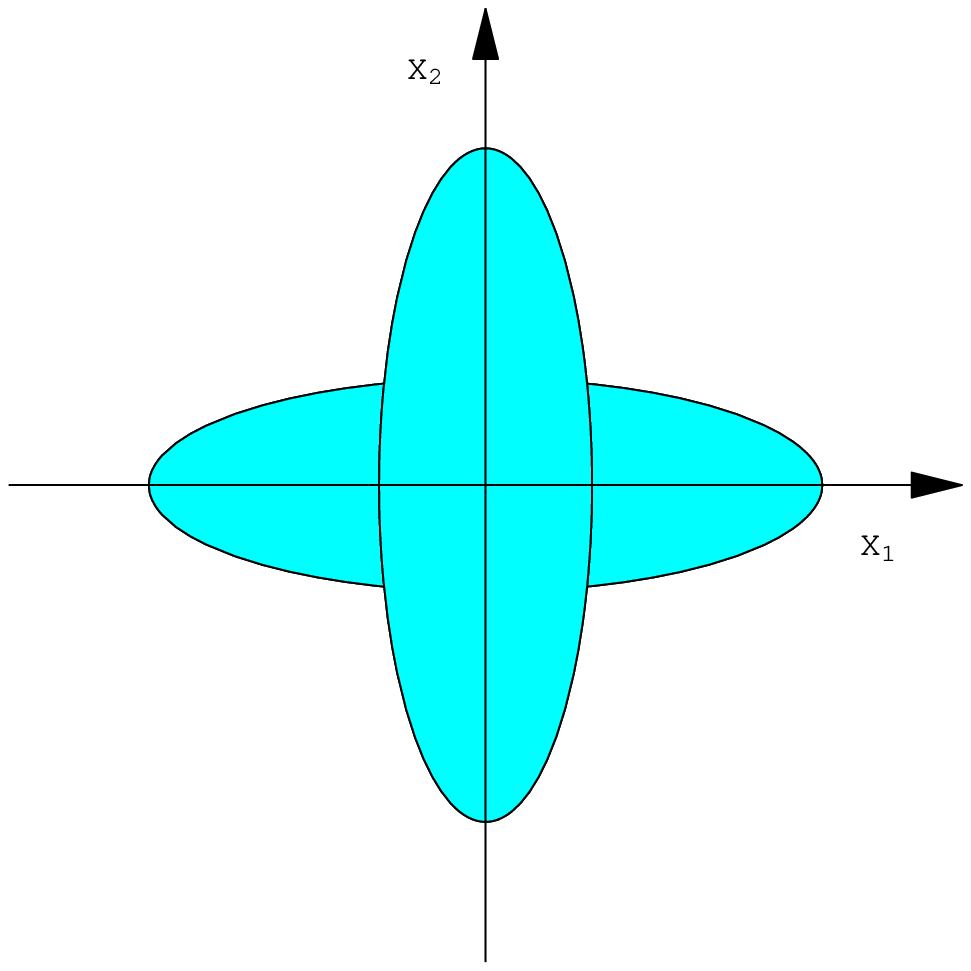}}
  \scalebox{0.4}{\includegraphics{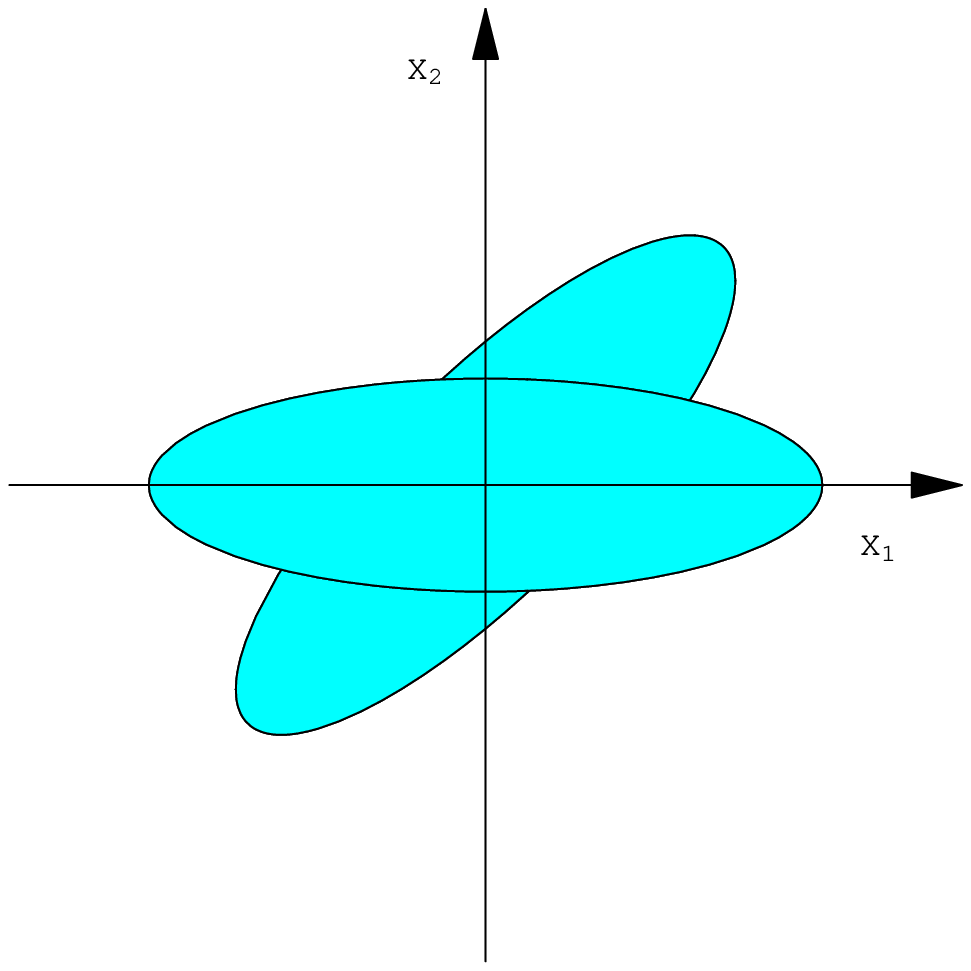}}
\caption{Beam splitter mixes two independently
  squeezed states, which error contours are rotated at different
  angles, and produces an entangled state. For input states rotated at
  the angles of $0$ and $\frac{\pi}{2}$ the outcome is EPR state ---
  left plot. For angles $0$ and $\frac{\pi}{4}$ the outcome is a
  general squeezed state --- right plot.}
\label{Fig:9_sec5}
\end{center}
\end{figure}

The rotated state preserves squeezing of the amplitude (phase)
quadrature if $\theta = \pm 2\lambda$, what corresponds to the
rotation at an angle $\pm\lambda$.  This rotation has a simple
geometrical picture of the beam splitter mixing of two independently
squeezed states, rotated at different angles, $0$ and $\lambda$. Such
mixing is producing an entangled state since the beam splitter is a
nonlocal operation on the two squeezed beams (Fig.~\ref{Fig:9_sec5}).

Let us investigate the HBT interference of the state
(\ref{eq:matrixC_sec5}). In contrast to the EPR case, the HBT
interference leads to phase-dependent in- and out-of-phase
visibilities
\begin{eqnarray}
\langle :I(\varphi_1) I(\varphi_2):\rangle &=& \frac{1}{2} (3n^2 + |m|^2)
\left[1+v_-\cos(\varphi_1 - \varphi_2)\right.\nonumber\\
&&\left.{}+ v_+ \cos(\varphi_1 + \varphi_2)\right],
\end{eqnarray}
\begin{equation}
v_{-} = \frac{n^2 + \frac{1}{2}(1-\cos 2\lambda)|m|^2}{3n^2 +
|m|^2}, \ \ v_{+} = \frac{\frac{1}{2}(1+\cos 2\lambda)|m|^2}{3n^2
+ |m|^2}\,.
\end{equation}
For these visibilities the set of  classical inequalities is valid
\begin{equation}
\frac{1}{4} \le v_{-} \le \frac{1}{2}, \; \; 0 \le v_{+} \le
\frac{1}{4}, \; \; v_{-} + v_{+} \le \frac{1}{2}\,.
\label{eq:classical_inequalities_sec5}
\end{equation}
Similar inequalities were derived in Sec. III for classical GSS.

This correlation function has been experimentally
demonstrated in \cite{Zhang} on Fig.[11], and has been used as an
experimental test for the quality of the EPR state. The reduction
of noise in the beam, below the vacuum level, was dependent on the
value of the mutual phase differences between two squeezed beams,
$\lambda$.

For input states from the borderline $|m|=n$ between quantum and
classical states we have the visibilities, $ v_{-} =
\frac{1}{8}(3-\cos2\lambda)$ and $v_{+} =
\frac{1}{8}(1+\cos2\lambda)$, that are clearly classical.

\begin{figure}
\begin{center}
\scalebox{0.8}{\includegraphics{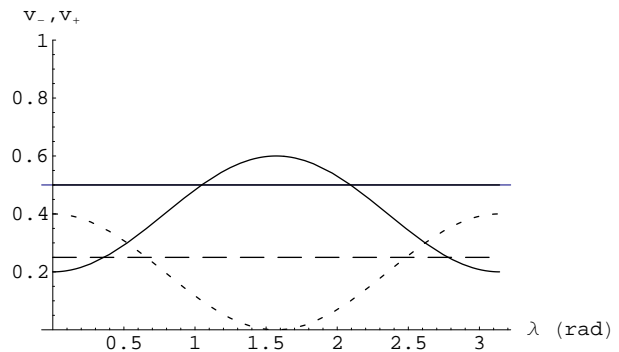}}
\caption{The values of $v_-$ (the solid curve) and $v_+$ (the dotted
  curve) for pure state with $n=1$ and $|m|=2$. For some angles
  $\lambda$, values of $v_-$ can exceed $\frac{1}{2}$ (the solid grid
  line) and values of $v_+$ can exceed $\frac{1}{4}$ (the dashed grid
  line). For these angles classical inequalities
  (\ref{eq:classical_inequalities_sec5}) are violated.}
\label{Fig:10_sec5}
\end{center}
\end{figure}

In Fig.~\ref{Fig:10_sec5} we have depicted the values of $v_-$ and
$v_+$ for a pure state with mean number of photons $n=1$ and $|m|=2$,
depending on phase $\lambda$. For some angles $\lambda$ the outcome is
entangled, values of its $v_-$ exceed $\frac{1}{2}$ or drops below
$\frac{1}{4}$ and values of its $v_+$ exceed $\frac{1}{4}$.

\section{More Complicated Case}

Let us suppose that two modes of correlated Gaussian states
investigated in the previous section are not squeezed equally and the
two squeezing parameters are not related to the correlation parameter
$m_c$ in any way. In such case we can have
\begin{eqnarray}
\langle a^{\dagger} a \rangle &=& \langle b^{\dagger} b \rangle =n, \;
\langle ab \rangle = -m_c, \nonumber \\
\langle a a\rangle &=& -me^{i \lambda_1}, \; \langle bb \rangle =
-me^{i\lambda_2},
\end{eqnarray}
where $\lambda_1+\lambda_2 \not= \pi$, and $m$ and $m_c$ are real for
simplicity. We assume the same mean number of photons $n$ in each
mode.

This state is separable if the following separability condition is
obeyed
\begin{equation}
m^2 + m_c^2 + |m_c| \sqrt{1+2(1+\cos\big(\lambda_1 +
  \lambda_2)\big)m^2}\le n(n+1).
\end{equation}

In this case, the in- and out-of-phase terms are not the only terms
that can contribute to the second-order interference. Few more cosine
terms appear due to the fact that the modes $a$ and $b$ are correlated
and these correlations do not cancel when evaluating the expressions
$\langle b^{\dagger}:(I_a+I_b):a\rangle$ and $\langle
a^{\dagger}:(I_a+I_b):b\rangle$ (\ref{eq:HBT_correlation_sec3}).

As a result we obtain the following second-order correlation
function
\begin{eqnarray}
\langle :I(\varphi_1) I(\varphi_2):
\rangle &\propto& 1+ v_{-} \cos{(\varphi_1-\varphi_2)} \nonumber \\
&& {} + v_{+} \cos{(\varphi_1+\varphi_2- \lambda_1 + \lambda_2})\nonumber\\
&& {} + v_{m} \big( \cos(\varphi_1-\lambda_1)+\cos(\varphi_2 + \lambda_2)
\nonumber\\
&& {} + \cos(\varphi_1 + \lambda_2) + \cos(\varphi_2 -
\lambda_1) \big) \nonumber
\end{eqnarray}
with three visibilities
\begin{eqnarray}
v_{-} = \frac{n^2+m_c^2}{3n^2+m_c^2+m^2}, && v_{+} =
\frac{m^2}{3n^2+m_c^2+m^2}, \nonumber \\
v_{m} &=& \frac{mm_c}{3n^2+m_c^2+m^2}.
\label{eq:visibilities_sec6}
\end{eqnarray}

The terms modulated by the visibility $v_{m}$ are genuine to
correlated squeezed states.

The appearance of $\lambda_1$ and $\lambda_2$ in the cosine terms is
due to the fact that these two beams are not squeezed equally. In this
case they do not give any contribute to the visibilities because the
absolute values of squeezing parameters are $\lambda_1$- and
$\lambda_2$-independent.

\begin{figure}[h!]  \begin{center}
\scalebox{0.8}{\includegraphics{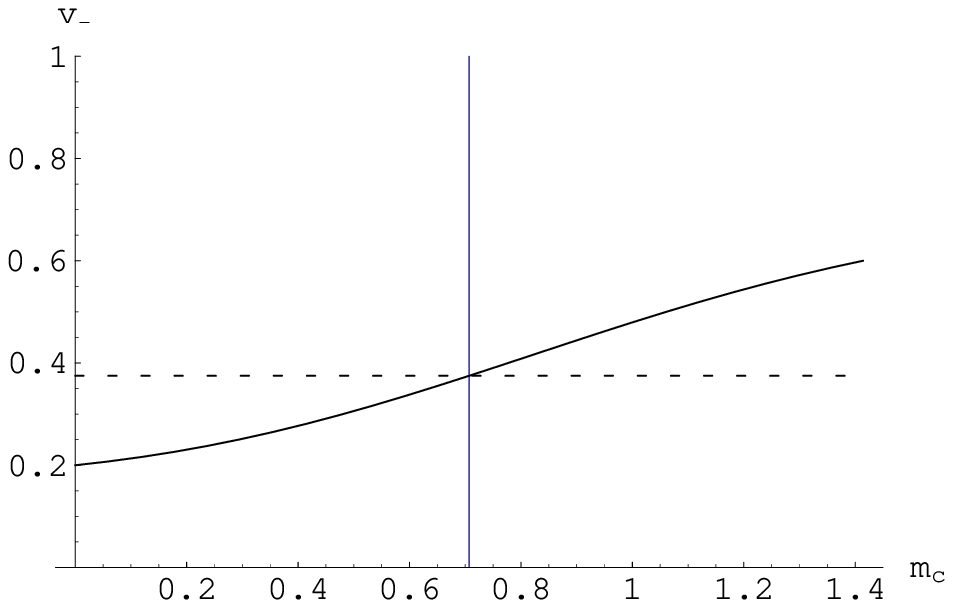}}
\scalebox{0.8}{\includegraphics{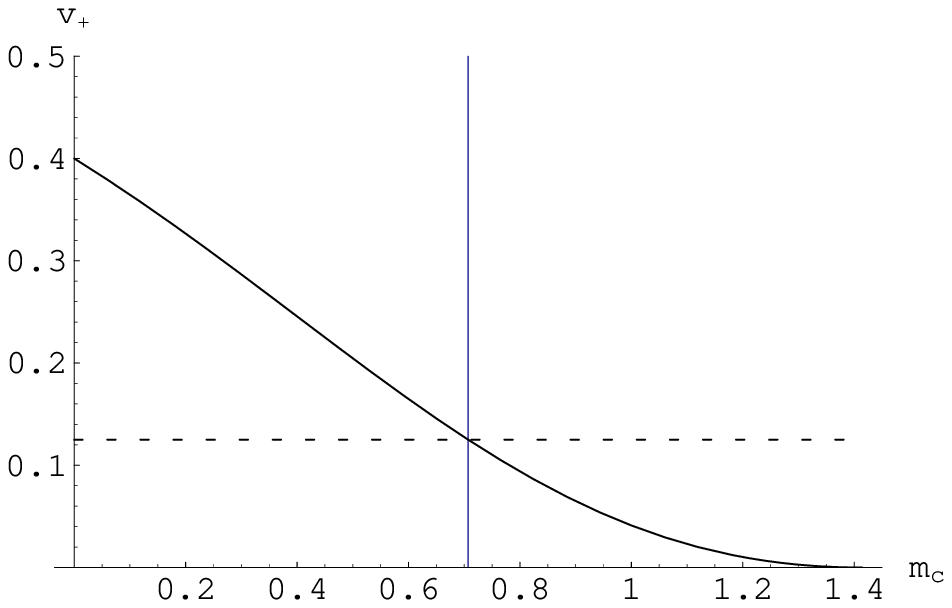}}
\scalebox{0.8}{\includegraphics{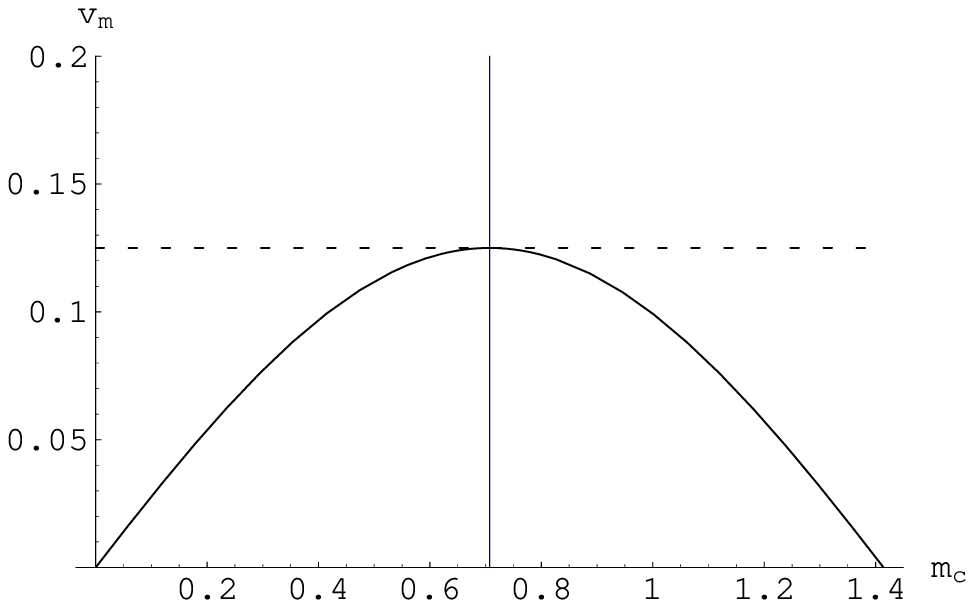}}
\caption{The visibilities (\ref{eq:visibilities_sec6}) evaluated
  for $\lambda_1=0$, $\lambda_2=\frac{\pi}{2}$ with $n=1$ and $m =
  \sqrt{2}-m_c$.  The values of $m_c >\frac{1}{\sqrt{2}}$ --- the
  vertical grid line, correspond to entangled state. $\sqrt{2}$ is the
  maximal value obeying positivity condition. The values of $v_-
  >\frac{3}{8}$ and $v_+ < \frac{1}{8}$ are genuine to entangled
  state. Values of $v_m$ are bounded from above by $\frac{1}{8}$.}
\label{Fig:11_sec6}
\end{center}
\end{figure}

In Fig.~\ref{Fig:11_sec6} visibilities $v_-$, $v_+$, $v_m$ are
depicted for a state with a mean number of photons equal to $n=1$,
squeezing parameter $m = \sqrt{2}-m_c$ and phases $\lambda_1=0$,
$\lambda_2=\frac{\pi}{2}$.  The border value of $m_c$ between
separable and entangled states is equal to $\frac{1}{\sqrt{2}}$. The
values of $v_-$ greater than $\frac{3}{8}$ and $v_+$ less than
$\frac{1}{8}$ correspond to entangled states. For $v_m$ the
monotonicity decides if the state is entangled.

\section{Conclusions}

In this paper we have presented a new method for witnessing quantum
squeezing and entanglement of one- and two-mode mixed Gaussian states.
This method relies on HBT interference and second-order intensity
correlations. For single mode GSS, a quantum witness operator has been
introduced in order to distinguish classical and quantum squeezed
states.  The quantum features of HBT interference can be probed with
the help of HBT witness operator. We have shown that for EPR states
this witness operator provides a good test of quantum separability. We
have applied the HBT interference and the HBT witness operator to
discuss quantum properties of entangled squeezed states generated by a
beam splitter. We have shown that the HBT witness operators can be
related to second-order visibilities and various classical
inequalities.

\begin{acknowledgments}
  This work was partially supported by a KBN grant No. 2PO3B 02123.
\end{acknowledgments}

\end{document}